# High-resolution low-coherence Brillouin optical correlation-domain reflectometry with suppressed systematic error

Kenta Otsubo,[1] Takaki Kiyozumi,[1] Kohei Noda,[1,2,3]
Kentaro Nakamura,[2] Heeyoung Lee,[4] and Yosuke Mizuno[1]

[1] *Faculty of Engineering, Yokohama National University, 79-5 Tokiwadai, Hodogaya-ku, Yokohama 240-8501, Japan*

[2] *Institute of Innovative Research, Tokyo Institute of Technology, 4259, Nagatsuta-cho, Midori-ku, Yokohama 226-8503, Japan*

[3] *Graduate School of Engineering, The University of Tokyo, 7-3-1 Hongo, Bunkyo-ku, Tokyo 113-8656, Japan*

[4] *Graduate School of Engineering and Science, Shibaura Institute of Technology, 3-7-5 Toyosu, Koto-ku, Tokyo 135-8548, Japan*

*Author e-mail addresses: otsubo-kenta-wv@ynu.jp, mizuno-yosuke-rg@ynu.ac.jp*

**Abstract:** We show that the systematic error unique to Brillouin optical correlation-domain reflectometry (BOCDR) can be effectively suppressed by use of low-coherence light, and demonstrate distributed strain measurement with ~3 cm spatial resolution.

## 1. Introduction

Brillouin scattering in optical fibers has been extensively used to develop distributed sensing systems, because the Brillouin frequency shift (BFS) depends on applied strain and ambient temperature [1]. From the point of view of spatially resolving method, time- [2–4], frequency- [5,6], and correlation-domain [7–12] techniques have been implemented. Each technique has its own advantages and disadvantages, as summarized in [13–15], and here we focus on correlation-domain techniques, which have relatively high spatial resolution and random accessibility (ability to measure strain and temperature at arbitrary points in a sensing fiber at high speed).

Brillouin-based distributed correlation-domain sensing techniques are divided into two groups: Brillouin optical correlation-domain analysis (BOCDA) [8,9] and Brillouin optical correlation-domain reflectometry (BOCDR) [10–12]. BOCDA, which requires light injection into both ends of a sensing fiber, is based on stimulated Brillouin scattering (SBS), which leads to its high signal-to-noise ratio (SNR) and high spatial resolution, at the cost of a relatively complicated setup including an electro-optic modulator (a single-sideband modulator is often used to generate frequency-shifted probe light) and a lock-in amplifier. Another disadvantage of BOCDA is that if the sensing fiber is broken during operation, the measurement cannot be performed any longer. BOCDR, on the other hand, requires light injection into only one end of a sensing fiber because it is based on spontaneous Brillouin scattering. Compared to BOCDA, the SNR and the achievable spatial resolution of BOCDR are low. However, its one-end-accessibility provides convenience in embedding the sensors in structures, and the setup can be relatively simple and low-cost. In addition, even if the sensing fiber is broken during operation, the measurement can continue to the point of breakage.

In the past, to improve the spatial resolution of BOCDA, researchers have developed low-coherence BOCDA, in which a low-coherence light source is used and its coherence length determines the spatial resolution. Extremely high spatial resolution can be achieved, but the measurement position needs to be scanned along the sensing fiber by adjusting the variable delay line, resulting in low operation speed. Hotate et al. [16] first demonstrated distributed strain measurement by low-coherence BOCDA using a noise-modulated laser. Subsequently, Cohen et al. [17] demonstrated the operation using much broader amplified spontaneous emission (ASE) of an erbium-doped fiber as a light source and achieved 4 mm resolution. Zafari et al [18] used ASE-based low-coherence BOCDA to measure the BFS distribution along a chalcogenide waveguide with 800 μm resolution. Wang et al. [19,20] demonstrated low-coherence BOCDA using a chaotic laser and achieved a resolution of 3.5 mm. Thus, low-coherence BOCDA has been widely studied to achieve high spatial resolution, but its two-end-access nature and relatively high cost due to probe light generation and lock-in detection remain unsolved. Considering the relatively low SNR of BOCDR, such high resolutions are not expected, but it is of great interest to experimentally demonstrate the operation of low-coherence BOCDR with single-end accessibility.

Recently, Zhang et al. [21] have developed low-coherence BOCDR (referred to as noise-correlated Brillouin optical reflectometry in the literature), and distributed BFS measurement with a spatial resolution of 19 cm has been demonstrated. The sensing range was relatively long (~250 m), but the measurement was performed every 30 cm along the FUT due to the minimal delay step size of an optical delay generator. Although the dependences of the spatial resolution on some experimental parameters were investigated, the BFS measurement accuracy was not compared with that of standard BOCDR using a sinusoidally frequency-modulated laser.

In this work, by using low-coherence BOCDR with a noise-modulated laser, we perform distributed BFS measurements with a spatial resolution of ~3 cm and show the systematic error unique to BOCDR can be effectively suppressed by this configuration. We detected strains ranging from 0 to 1% at intervals of 0.2%, demonstrating the practicality of the system.

## 2. Principles

In standard BOCDR, the laser output is sinusoidally frequency-modulated using a technique called the synthesis of optical coherence functions [7], and correlation peaks (which act as sensing points) are generated along a sensing fiber [11]. We often limit the length of the sensing fiber so that only a single correlation peak exists in the fiber. By sweeping the modulation frequency, the position of the correlation peak can be scanned along the fiber, allowing for distributed BFS measurements. The spatial resolution is determined by the modulation frequency and modulation amplitude [10].

In contrast, low-coherence BOCDR uses a low-coherence light source and generates a single correlation peak (0th order) at the point where there is no optical-path-length difference (between the signal and reference paths). By changing the optical path length using a variable optical delay line, the sensing point can be scanned along the fiber for distributed measurement. The spatial resolution is determined by the coherence length of the light source [22], as

$$\Delta z = \frac{2 ln 2}{\pi} \cdot \frac{c}{n \Delta f}, \qquad (1)$$

where $c$ is the speed of light in vacuum, $n$ is the refractive index of the fiber core, and $\Delta f$ is the linewidth of the light source. Thus, although standard BOCDR and low-coherence BOCDR are both based on spontaneous Brillouin scattering, the spatial resolution and the measurement range are determined by different factors.

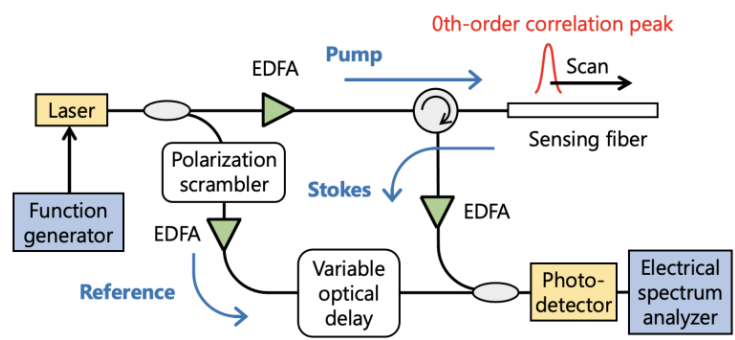

Fig. 1. Experimental setup of low-coherence BOCDR. EDFA: erbium-doped fiber amplifier.

Figure 1 shows the experimental setup of low-coherence BOCDR. Considering that the signal-to-noise ratio of BOCDR is lower than that of BOCDA, we used a noise-modulated distributed-feedback laser diode (LD; 1551 nm; WSLS-934010C1124-58, CivilLaser), the linewidth of which can be controlled by the modulation amplitude, as the low-coherence light source instead of ASE or other types of broadband sources. The output of the LD was split into signal light and reference light by an optical coupler. The signal light was amplified to 25.0 dBm by an erbium-doped fiber amplifier (EDFA) and injected into a sensing fiber via an optical circulator. The spontaneous-Brillouin-scattered light from the sensing fiber was amplified to 1.8 dBm with another EDFA. The reference light was polarization-scrambled, amplified to 2.8 dBm, passed through a variable optical delay line (variable fiber length: 75 cm; ADL-200-25-SM-FS, Alnair Labs), and mixed with the returned signal light for self-heterodyne detection. Their beat signal was converted to an electrical signal with a photodetector, and then the Brillouin gain spectrum (BGS) was observed with an electrical spectrum analyzer (ESA). Distributed measurement can be performed by continuously changing the length of the variable optical delay line to sweep the sensing position along the sensing fiber.

## 3. Experiments

First, the laser linewidth was measured as a function of the peak-to-peak voltage of the noise modulation, as shown in Fig. 2(a). The direct current bias was 7.0 V. Due to the limited frequency resolution of an optical spectrum analyzer (AQ6370, Yokogawa Electric), the laser linewidth was measured directly only when the modulation voltage was in the range of 2.5–6.0 Vp-p. The laser linewidth without modulation (~3 MHz; described in a specification sheet) was also plotted. With increasing modulation voltage, the laser linewidth increased, but the dependence coefficient gradually decreased. These plots were then fitted with a second-order polynomial curve. Figure 2(b) shows the theoretical spatial resolution calculated from the fitted curve and Eq. (1). With increasing modulation voltage, the theoretical spatial resolution gradually became higher (the value became smaller) and was almost constant around 3 cm for higher modulation voltage.

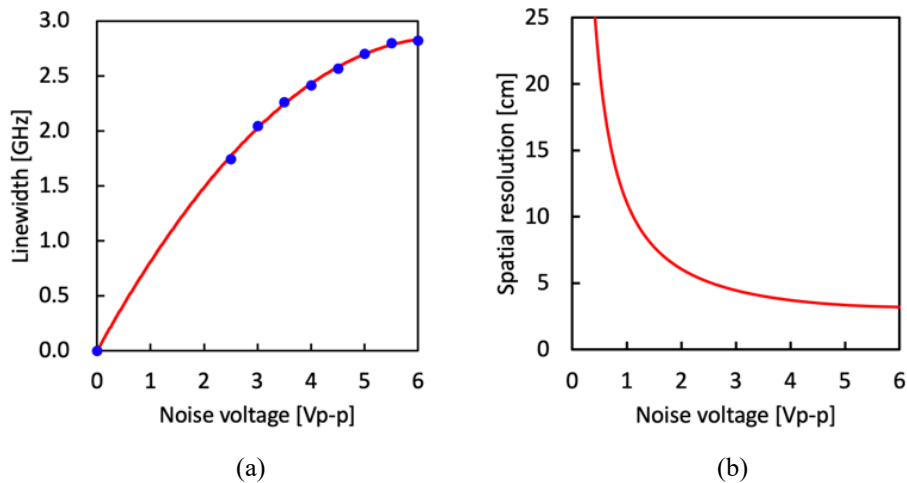

(a) (b)

Fig. 2. (a) Laser linewidth plotted as a function of the noise modulation voltage. The red solid curve is a polynomial fit. (b) Dependence of the theoretical spatial resolution on the noise modulation voltage.

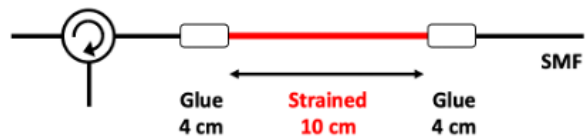

Fig. 3. Structure of the fiber under test.

Subsequently, distributed strain measurements based on low-coherence BOCDR were demonstrated. We employed a sensing fiber with the structure shown in Fig. 3. Strains up to 1.0% were applied to a 10-cm-long section of a 4.4-m-long silica single-mode fiber (SMF; BFS ~10.8 GHz at room temperature), which was fixed on movable stages by two 4-cm-long glued parts; the total fixed length was 18 cm. The BGS and BFS distributions (50 points) were measured along a 37.5-cm-long fiber section including the strained part. The noise modulation voltage was 6.0 Vp-p, which corresponds to a spatial resolution of 3.2 cm as shown in Fig. 2(b).

The normalized BGS distribution (measured when strain was 1.0%) and the BFS distributions measured for different applied strains are shown in Figs. 4(a) and 4(b), respectively. In both figures, the strained region was clearly identified. With increasing applied strain, the BFS linearly shifted to a higher frequency with a coefficient of 357 MHz/%, which is slightly lower than the typical value for silica SMFs. The gradual shift of the BFS reflects the strain distribution inside the glued parts (as shown in the following measurement). The fact that the strain was not concentrated on the strained part is the reason for the lower strain coefficient of the BFS mentioned above. What is striking here is that the BFS distributions are highly symmetric in the fiber longitudinal direction, showing that the systematic error, which is caused by the phase delay between amplitude and frequency modulations in a directly modulated laser in standard sinusoidal-modulation BOCDR [23,24], can be effectively suppressed by this configuration.

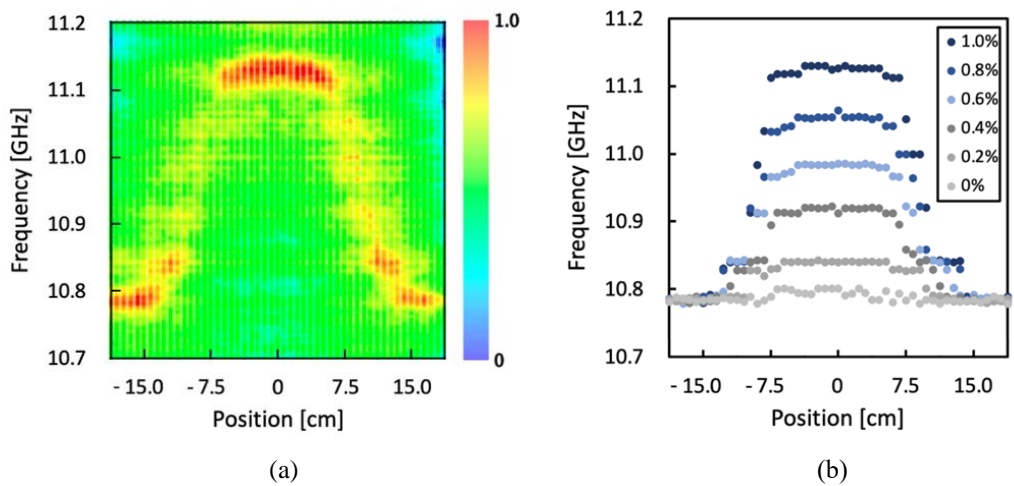

(a) (b)

Fig. 4. (a) Normalized BGS distribution measured when applied strain was 1.0%. (b) Measured BFS distributions with different applied strains.

## 4. Conclusions

On the basis of low-coherence BOCDR using a noise-modulated LD, we performed distributed BFS measurements with a spatial resolution of ~3 cm, which is ~6 times better than the previous report [21]. We also show the potential of this configuration for suppressing the systematic error unique to BOCDR effectively. We expect that in the future, low-coherence BOCDR will be one of the promising options for distributed strain sensing with its single-end accessibility, relatively high spatial resolution, system simplicity, cost efficiency, and improved measurement accuracy.